\pdfoutput=1

\documentclass[11pt]{article}

\usepackage[preprint]{acl}

\usepackage{times}
\usepackage{tabularx}
\usepackage{enumitem}
\usepackage{latexsym}
\usepackage{booktabs}
\usepackage[T1]{fontenc}

\usepackage[utf8]{inputenc}

\usepackage{microtype}

\usepackage{inconsolata}

\usepackage{graphicx}

\title{eSapiens: A Real-World NLP Framework for Multimodal Document Understanding and Enterprise Knowledge Processing}

\author{
Isaac Shi \quad
Zeyuan Li \quad
Wenli Wang \quad
Lewei He \quad
Yang Yang \quad
Tianyu Shi \\
\textbf{eSapiens Team}
}

\begin{document}
\maketitle
\begin{abstract}
We introduce eSapiens, a unified question-answering system designed for enterprise settings, which bridges structured databases and unstructured textual corpora via a dual-module architecture. The system combines a Text-to-SQL planner with a hybrid Retrieval-Augmented Generation (RAG) pipeline, enabling natural language access to both relational data and free-form documents. To enhance answer faithfulness, the RAG module integrates dense and sparse retrieval, commercial reranking, and a citation verification loop that ensures grounding consistency. We evaluate eSapiens on the RAGTruth benchmark across five leading large language models (LLMs), analyzing performance across key dimensions such as completeness, hallucination, and context utilization. Results demonstrate that eSapiens outperforms a FAISS baseline in contextual relevance and generation quality, with optional strict-grounding controls for high-stakes scenarios. This work provides a deployable framework for robust, citation-aware question answering in real-world enterprise applications.
\end{abstract}

\section*{Introduction}

In real-world enterprise scenarios, the demand for large-scale, unified language systems capable of handling heterogeneous data formats—such as structured relational databases and unstructured long-form documents—continues to grow. Despite the advancements of large language models (LLMs) in open-domain tasks, their direct deployment in business-critical environments remains challenging due to domain-specific requirements, tool integration needs, latency constraints, and issues of interpretability.

To address these challenges, we introduce {eSapiens, an industrial-grade language intelligence platform engineered to support end-to-end human-AI collaboration across both structured and unstructured data pipelines. The system combines two complementary capabilities: a Text-to-SQL (T2S) module for translating natural language questions into executable queries over structured databases, and a Retrieval-Augmented Generation (RAG) module for retrieving and synthesizing information from long-form, domain-specific documents. These two components are orchestrated in a unified framework that supports plug-and-play integration into enterprise workflows, enabling both semantic querying and contextualized generation under real-time constraints.

Our work draws inspiration from prior research in semantic parsing over tabular data and retrieval-augmented question answering. However, most existing systems either focus on a single data modality or lack the robustness required for deployment in production-grade environments. eSapiens extends this line of work by integrating task-specific execution logic, domain adaptation, and conversational memory into a unified, multi-tenant platform suitable for real-world enterprise use.

The contributions of this paper are as follows:

\begin{itemize}
\item We present the design and implementation of eSapiens, a production-grade language intelligence system that unifies Text-to-SQL and retrieval-augmented generation (RAG) pipelines for natural language interaction with structured and unstructured enterprise data.

\item We detail the system's modular architecture, which supports flexible integration with heterogeneous data sources and enterprise infrastructures, including multi-user, multi-agent orchestration and real-time query execution.

\item We report empirical results and deployment case studies that demonstrate the effectiveness, robustness, and practical utility of the system in business-critical scenarios.
\end{itemize}

\begin{figure*}[t]
  \centering
  \includegraphics[width=\textwidth]{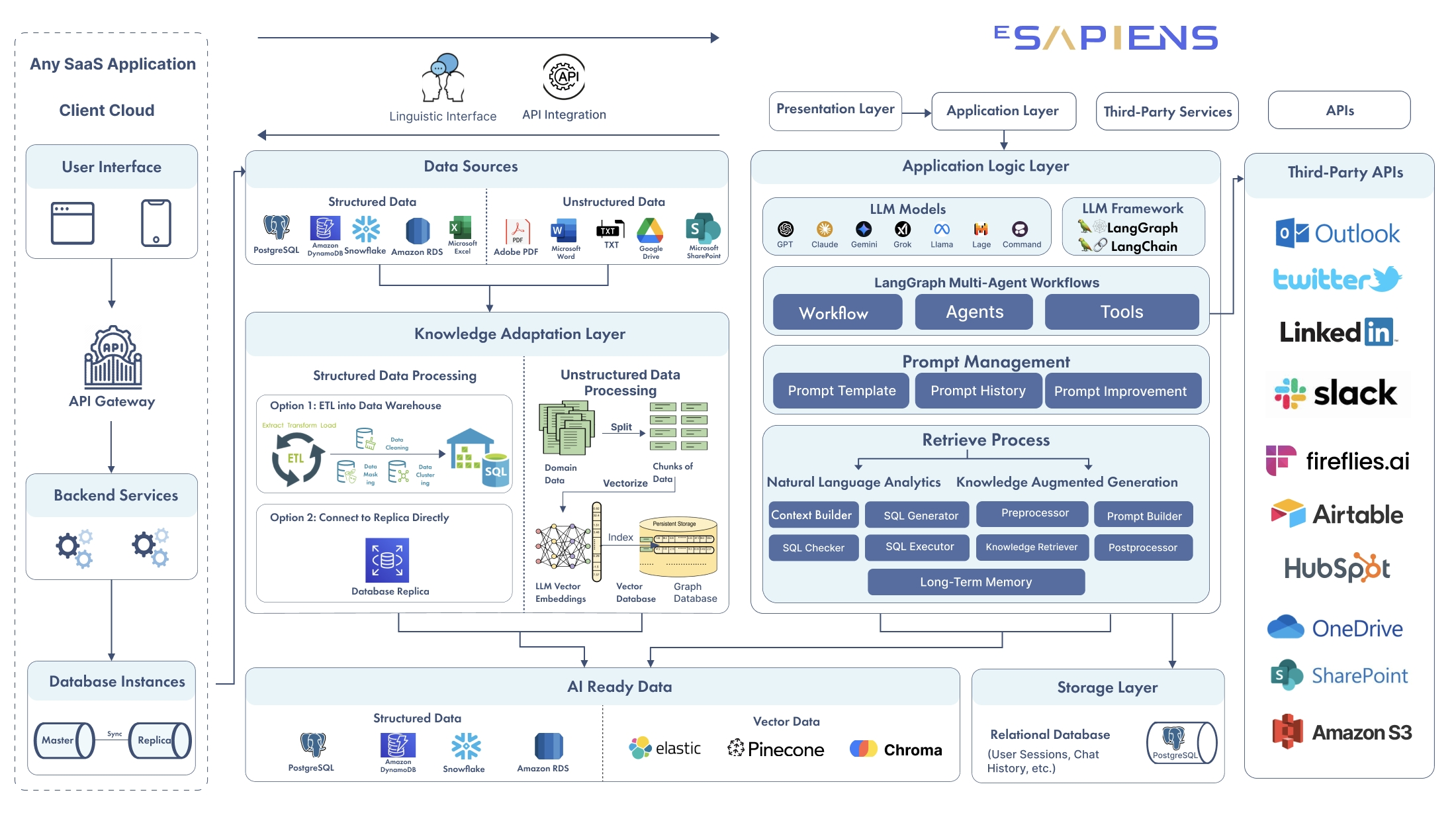}
  \caption{Overall architecture of the eSapiens platform.}
  \label{fig:system_architecture}
\end{figure*}

\section*{Related Work}

\subsection*{Text-to-SQL}
Text-to-SQL (T2S) technology translates natural language questions into executable SQL queries. Early neural approaches established foundational methods: Seq2SQL~\cite{zhong_seq2sql_2017} pioneered sequence-to-sequence architectures for direct SQL generation, while SQLNet~\cite{xu_sqlnet_2017} introduced sketch-based decoding to improve structural robustness. Subsequent innovations addressed complex query handling; SyntaxSQLNet~\cite{yu_syntaxsqlnet_2018} incorporated syntactic trees for nested queries, and IRNet~\cite{guo_irnet_2019} leveraged intermediate representations to resolve schema linking ambiguities. The field advanced significantly with the Spider benchmark~\cite{yu_spider_2018}, which established rigorous cross-domain evaluation across 200 databases. This catalyzed architectures like RAT-SQL~\cite{wang_rat_2020} that employed relation-aware transformers to model database dependencies. Recent large language model (LLM) integrations demonstrate state-of-the-art performance: DIN-SQL~\cite{pourreza_din_2023} uses decomposition prompting for multi-step queries, DAIL-SQL~\cite{gao2023texttosq} achieves 86.6\% execution accuracy via optimized prompt engineering, and C3~\cite{dong_c3_2023} enforces zero-shot cross-domain consistency. Despite these advances, production deployment faces persistent challenges in interpretability and concurrent user support—gaps our modular architecture specifically addresses.

\subsection*{Retrieval-Augmented Generation}
Retrieval-Augmented Generation (RAG) enhances language models through dynamic knowledge retrieval. Foundational frameworks include REALM~\cite{guu_realm_2020}, which integrated retrieval during pre-training, and RAG~\cite{lewis_rag_2020}, which established end-to-end trainable retrieval-generation pipelines. Dense retrieval techniques markedly improved precision; DPR~\cite{karpukhin_dpr_2020} enabled fine-grained passage matching via dual-encoders, while ColBERT~\cite{khattab_colbert_2020} balanced accuracy and efficiency through late interaction mechanisms. Recent enterprise-focused extensions tackle scalability challenges: Atlas~\cite{izacard_atlas_2022} demonstrated few-shot adaptation for knowledge-intensive tasks, and KILT~\cite{petroni_kilt_2021} standardized evaluation across domain-specific corpora. Nevertheless, real-world deployment struggles with heterogeneous document processing, context window limitations, and latency constraints—issues our multi-channel indexing and domain-tuned rerankers directly mitigate.

\section*{System Architecture}

\subsection*{Overview}

The eSapiens system is designed as a unified platform for natural language interaction with both structured and unstructured data. Figure~\ref{fig:system_architecture} illustrates the overall architecture, which consists of five key layers: the user interface layer, orchestration and memory layer, tool integration layer, core reasoning modules, and knowledge base layer.

The top-level user interface layer provides web-based endpoints for human-AI collaboration, supporting multiple users and roles. User inputs are parsed and passed to the orchestration layer, which manages session memory, determines routing strategies, and invokes appropriate tools or reasoning modules.

The core of the system lies in two complementary reasoning modules: a Text-to-SQL (T2S) component for handling structured database queries, and a Retrieval-Augmented Generation (RAG) component for processing unstructured documents. These modules are designed to work independently or cooperatively, depending on the query type.

The tool integration layer encapsulates third-party systems such as data sources, vector databases, large language models (LLMs), and internal plugins. It supports both synchronous and asynchronous calls, enabling real-time and batch-style execution.

At the bottom, the knowledge base layer supports heterogeneous knowledge sources, including relational databases, enterprise documents, vector indexes, and structured metadata. This layer provides the foundation for semantic reasoning, search, and generation.

\subsection*{RAG Module Architecture}

The RAG module in eSapiens is designed as a unified, modular workflow that transforms user queries into grounded, fully-cited answers across multimodal and enterprise-scale sources. It is implemented using the LangGraph framework, which coordinates a series of functional agents to process each query through a dynamic and stateful execution graph.

Each query begins with the initialization of a state object that maintains the dialogue context, retrieved content, plugin results, visual elements, and citation traces. A central supervisor agent analyzes the query intent and decides the appropriate path among document retrieval, SQL querying, external plugin calls, chart rendering, or image analysis.

For internal knowledge retrieval, the system uses a hybrid search index combining dense vector similarity (via HNSW) and sparse keyword search (via BM25) powered by Elasticsearch. Documents are ingested from various formats such as DOCX, PDF, and TXT, segmented into overlapping windows of approximately 1000 tokens, and embedded using a large-scale embedding model. The retrieval pipeline first collects the top 200 candidate passages, reranks them using a commercial reranking model, and selects the top 50 snippets for response generation.

All generated answers are citation-grounded. Each sentence in the draft is verified against the retrieved snippets. If the verification fails, the system automatically triggers regeneration until all cited statements are supported. This ensures factual integrity and traceability.

When internal retrieval proves insufficient, the system invokes additional agents. Web search is conducted through an external search API. Plugins registered within the platform can be called to fetch structured results from external services. Visual or chart-related queries activate agents for image parsing and chart synthesis. Finally, an answer optimization agent assembles the result, integrates intermediate outputs, and streams the response to the user in real time using server-sent events.

This architecture enables reliable, extensible, and interactive enterprise retrieval-augmented generation. By coordinating multimodal tools and enforcing citation verification, the system ensures both responsiveness and answer fidelity.

\begin{table*}[t]
\centering
\caption{Performance of eSapiens (Chunk = 500): Recall and Precision at Different $k$}
\label{tab:chunk500}
\renewcommand{\arraystretch}{1.1}
\setlength{\tabcolsep}{4pt}
\resizebox{\textwidth}{!}{
\begin{tabular}{lcccccc|cccccc}
\toprule
\textbf{Dataset} & \multicolumn{6}{c|}{\textbf{Recall@k (\%)}} & \multicolumn{6}{c}{\textbf{Precision@k (\%)}} \\
\cmidrule(lr){2-7} \cmidrule(lr){8-13}
 & k=1 & k=2 & k=4 & k=8 & k=16 & k=50 & k=1 & k=2 & k=4 & k=8 & k=16 & k=50 \\
\midrule
PrivacyQA   & 18.15 & 25.87 & 49.28 & 64.07 & 85.63 & 96.47 & 18.50 & 14.02 & 13.18 & 9.26 & 4.74 & 5.28 \\
ContractNLI & 4.91  & 9.33  & 16.09 & 25.83 & 35.04 & 46.90 & 5.08  & 5.59  & 5.04  & 3.67 & 2.52 & 1.75 \\
MAUD        & 0.52  & 2.48  & 4.39  & 7.24  & 14.03 & 22.60 & 1.94  & 2.63  & 2.05  & 1.77 & 1.79 & 1.12 \\
CUAD        & 3.17  & 7.33  & 18.26 & 28.67 & 42.50 & 55.66 & 3.53  & 4.18  & 6.18  & 5.06 & 3.93 & 2.74 \\
ALL         & 7.26  & 11.52 & 20.40 & 27.94 & 41.37 & 51.82 & 7.49  & 6.82  & 6.65  & 5.02 & 3.78 & 2.68 \\
\bottomrule
\end{tabular}
}
\end{table*}

\subsection*{T2S Module Architecture}
The T2S (Text-to-SQL) module in eSapiens serves as a bridge between natural language instructions and structured enterprise data, enabling users to query databases without prior SQL knowledge. The module is built on a LangGraph-based workflow that coordinates SQL generation, execution, error correction, and answer formatting in a robust, modular pipeline.

Each query initializes a shared state containing the user question, chat history, metadata, and execution trace. A dedicated Text2SQL agent generates SQL statements by prompting an LLM with schema context, sample rows, and the user instruction. The system supports multiple SQL dialects and dynamically retrieves schema metadata from the connected data warehouse (e.g., BigQuery or MySQL).

Generated SQL is first validated through dry runs or direct execution. If the query fails due to syntax errors, missing entities, or returns empty results, the system triggers an error-aware retry mechanism. The failed SQL statement and error messages are injected into the prompt for a second-pass correction. The retry loop is bounded to avoid infinite cycles, and fallback strategies (e.g., returning tabular placeholders or suggesting query reformulations) ensure graceful degradation.

For valid results, the system parses and structures the output. If the query returns quantitative data, a chart generation agent automatically decides whether to visualize the results via bar, line, or pie charts, based on heuristics and user hints. If visualization is not feasible (e.g., non-numeric or sparse data), the system reverts to textual summarization with data highlights.

Finally, an answer optimizer agent formats the output with appropriate citations, explanation context, and semantic alignment with the original user instruction. The full response is streamed back to the user through server-sent events (SSE), ensuring responsiveness and interactivity throughout.

\begin{table*}[t]
\centering
\caption{Performance of eSapiens (Chunk = 1000): Recall and Precision at Different $k$}
\label{tab:chunk1000}
\renewcommand{\arraystretch}{1.1}
\setlength{\tabcolsep}{4pt}
\resizebox{\textwidth}{!}{
\begin{tabular}{lcccccc|cccccc}
\toprule
\textbf{Dataset} & \multicolumn{6}{c|}{\textbf{Recall@k (\%)}} & \multicolumn{6}{c}{\textbf{Precision@k (\%)}} \\
\cmidrule(lr){2-7} \cmidrule(lr){8-13}
 & k=1 & k=2 & k=4 & k=8 & k=16 & k=50 & k=1 & k=2 & k=4 & k=8 & k=16 & k=50 \\
\midrule
PrivacyQA   & 10.10 & 20.24 & 28.84 & 54.95 & 71.44 & 94.47 & 8.97  & 10.31 & 7.81  & 7.34  & 5.16 & 2.64 \\
ContractNLI & 4.81  & 8.72  & 12.62 & 17.85 & 25.54 & 39.78 & 2.28  & 2.47  & 1.84  & 1.33  & 0.89 & 0.42 \\
MAUD        & 0.52  & 2.48  & 3.05  & 4.57  & 7.31  & 13.60 & 1.33  & 1.07  & 0.84  & 0.64  & 0.53 & 0.32 \\
CUAD        & 3.62  & 10.47 & 20.63 & 32.46 & 45.24 & 62.30 & 2.12  & 3.08  & 3.17  & 2.70  & 2.01 & 0.96 \\
ALL         & 4.93  & 10.34 & 16.29 & 27.46 & 37.38 & 52.54 & 3.68  & 4.23  & 3.42  & 3.00  & 2.15 & 1.09 \\
\bottomrule
\end{tabular}
}
\end{table*}
\begin{table*}[htbp]
\caption{Evaluation results of RAG models on RAGtruth}
\centering
\label{tab:trace_results}
\resizebox{\textwidth}{!}{%
\begin{tabular}{lcccccc}
\toprule
\textbf{Model} & \textbf{Completeness} & \textbf{Utilization} & \textbf{Context Relevance} & \textbf{hallucination} & \textbf{Accuracy} \\
\midrule
eSapiens-gpt4o & 0.4307 & 0.5224 & 0.3648 & 0.1823 & 3.85 \\
eSapiens-gpt4o-mini & 0.3433 & 0.4658 & 0.3785 & 0.2729 & 3.70 \\
eSapiens-claude-3.7 & 0.3840 & 0.4440 & 0.3648 & 0.1403 & 4.05 \\
eSapiens-gemini-1.5-pro & 0.4982 & 0.5179 & 0.3728 & 0.1712 & 4.0 \\
eSapiens-deepseek-r1 & 0.4571 & 0.5167 & 0.2581 & 0.1486 & 4.15 \\
faiss+top-2+short+gpt4o & 0.4450 & 0.4800 & 0.3294 & 0.0875 & 3.55 \\
faiss+top-2+short+gpt4o-mini & 0.4105 & 0.4467 & 0.3090 & 0.1524 & 3.15 \\
faiss+top-2+short+claude-3.7 & 0.4985 & 0.5102 & 0.3270 & 0.0860 & 3.75 \\
faiss+top-2+short+gemini-1.5-pro & 0.5346 & 0.5728 & 0.3296 & 0.1371 & 3.75 \\
faiss+top-2+short+deepseek-r1 & 0.5381 & 0.5102 & 0.3430 & 0.1139 & 4.15 \\
\bottomrule
\end{tabular}%
}

\end{table*}

\section*{Experiment}
The experimental evaluation of the RAG module in eSapiens is designed to analyze both the internal sensitivity of system configurations and the comparative effectiveness of retrieval strategies under citation-constrained generation. Our evaluation follows two directions.

We first conduct intra-system ablation studies to examine how different chunk sizes during document preprocessing and varying top-\emph{k} retrieval depths during reranking affect final answer quality. This analysis is performed on four legal-domain datasets from the LegalBench benchmark—PrivacyQA, CUAD, MAUD, and ContractNLI—covering a wide range of document styles, regulatory scopes, and question types.

We then perform cross-system retrieval comparisons to evaluate our hybrid pipeline—integrating dense HNSW-based vector search, BM25 lexical matching, and a commercial reranker—against a FAISS baseline. These comparisons are conducted on the RAGTruth benchmark, and results are measured across four key metrics: citation faithfulness, contextual relevance, factual correctness, and information completeness. The results collectively affirm that the design of the eSapiens RAG module ensures robust and generalizable performance across varied question answering tasks.

\subsection*{Retrieval Performance on Long-form Legal QA}
We evaluate the eSapiens RAG module across four long-form legal QA datasets—\textit{PrivacyQA}, \textit{CUAD}, \textit{MAUD}, and \textit{ContractNLI}—to examine how chunk size influences retrieval quality. As shown in Tables~\ref{tab:chunk500} and~\ref{tab:chunk1000}, several key trends emerge:

\begin{itemize}[leftmargin=*, itemsep=2pt]
  \item \textbf{Recall dominates:} Larger chunk windows (1000 tokens) consistently yield higher Recall@k across most datasets, especially at $k=50$, due to better semantic preservation.
  \item \textbf{PrivacyQA exception:} In Table~\ref{tab:chunk500}, chunk size 500 performs better at low $k$ (e.g., Recall@1: 18.15\% vs. 10.10\%) due to fragmented text structure.
  \item \textbf{Precision tradeoff:} Although smaller chunks slightly improve Precision@k, this is less critical in generation workflows where LLMs like GPT-4o can handle irrelevant context.
  \item \textbf{Practical recommendation:} Chunk size 1000 provides a better recall–completeness tradeoff and is better aligned with real-world usage in enterprise QA scenarios.
\end{itemize}

Overall, the results support the default design in eSapiens, which favors 1000-token chunks for robust and semantically complete retrieval.

\subsection*{TRACe Evaluation}

To further evaluate the response quality of eSapiens beyond retrieval effectiveness, we conduct a detailed comparative study using the TRACe benchmark. This framework enables fine-grained, multi-faceted assessment of generated outputs from large language models (LLMs) across five key dimensions: completeness, utilization, context relevance, hallucination, and accuracy. Our objective is to analyze how retrieval design—particularly the contrast between the eSapiens hybrid pipeline and a FAISS baseline—affects answer fidelity, grounding behavior, and user satisfaction.

The experiment spans five representative LLMs: GPT-4o, GPT-4o-mini, Claude 3.7, Gemini 1.5 Pro, and DeepSeek R1. For each, we evaluate both pipelines using a consistent and representative question set from the RAGTruth benchmark. Results are presented in Table~\ref{tab:trace_results}.

Across dimensions, several key findings emerge:

\noindent\textbf{Hallucination:} The FAISS-based pipeline exhibits consistently lower hallucination rates. This is attributed to its strict reliance on retrieved passages and refusal to generate unsupported answers. In contrast, eSapiens allows limited abstraction in generation, which—while enhancing fluency—may introduce minor unsupported claims. To mitigate this, eSapiens includes a ``strict grounding'' preset that enforces hard constraints (e.g., citation for every sentence, N/A fallback), making it suitable for compliance-critical scenarios.

\noindent\textbf{Completeness:} FAISS retrieval tends to yield higher completeness scores. Its top-$k$ retrieval structure ensures dense content overlap with expected answer spans. eSapiens, although more flexible, occasionally abstracts or restructures the answer, leading to partial omission of minor details.

\noindent\textbf{Context Relevance:} eSapiens significantly outperforms FAISS in aligning retrieved passages with user intent. The hybrid pipeline’s dense + sparse retrieval with reranking enables selection of more semantically targeted passages, which proves particularly effective when the question requires nuanced understanding.

\noindent\textbf{Utilization:} Models such as Gemini and GPT-4o show superior utilization of context within the eSapiens framework, reflecting their ability to reason over larger, well-ranked contexts. This suggests that retrieval quality and model capability must be jointly considered for optimal performance.

\noindent\textbf{Accuracy:} Human ratings indicate that eSapiens generations are generally more natural, coherent, and aligned with user expectations. This is largely due to post-processing components in eSapiens (e.g., formatting, rephrasing) that improve readability without compromising correctness.

\medskip
\noindent
In summary, the TRACe evaluation confirms that eSapiens achieves strong grounding and output quality by balancing citation fidelity, context richness, and linguistic fluency. While the FAISS baseline provides a safety-first alternative with stricter factuality guarantees, eSapiens offers a more robust and adaptable framework suitable for diverse real-world use cases.

\section*{Conclusion}

We present eSapiens, a unified natural language interface for enterprise data access that integrates a Text-to-SQL planner for structured queries and a hybrid RAG pipeline for unstructured content grounding. Through modular retrieval and citation verification, eSapiens achieves strong performance on RAGTruth, outperforming FAISS-only baselines across multiple LLMs in relevance, fluency, and factual grounding.

On the structured side, the T2S planner features rule-based fallback mechanisms to ensure robustness in noisy, real-world queries. Together, the two modules provide a flexible and faithful interface for diverse enterprise scenarios. Future work will explore tighter integration between RAG and T2S for mixed-schema reasoning and continuous improvement via user feedback.

In summary, eSapiens provides a robust and extensible solution for enterprise-level question answering, bridging structured and unstructured data with high fidelity and practical reliability.

\bibliography{custom}

\appendix

\section{T2S Evaluation \& Results}
\label{sec:appendix}

To gauge real-world robustness, we issued seven natural-language queries to eSapiens T2S and three popular text-to-SQL agents (anonymized here as Baseline Agents). The prompts span both logistics and retail schemas, including unit conversion, fuzzy text matching, and multi-step analytics. Table~\ref{tab:practice} summarizes the most common failure patterns observed in the baselines and how eSapiens addressed them.

\begin{itemize}[leftmargin=*,itemsep=2pt]
\item \emph{Self-healing pipeline}: The bounded retry loop salvaged several queries that baseline agents either failed to execute or returned empty results.
\item \emph{Semantic guardrails}: Heuristics for unit conversion, date range enforcement, and fuzzy text handling prevented “looks-right-but-wrong” answers from reaching users.
\item \emph{Readable insights}: When asked for actionable advice, eSapiens appended narrative explanations, whereas baseline agents returned raw tabular outputs.
\item \emph{Compliance by construction}: All queries remained read-only, and only eSapiens successfully rejected unauthorized column access.
\end{itemize}

\begin{table*}[h!]
\centering
\caption{Observed error patterns and eSapiens handling}
\label{tab:practice}
\renewcommand{\arraystretch}{1.15}
\setlength{\tabcolsep}{6pt}
\begin{tabularx}{\textwidth}{@{}
  >{\raggedright\arraybackslash}X
  >{\raggedright\arraybackslash}X
  >{\raggedright\arraybackslash}X
@{}}
\toprule
\textbf{Prompt Theme} &
\textbf{Typical Pitfall for Baseline Agents} &
\textbf{How eSapiens Handled It} \\
\midrule
Non-existent status codes &
Agent-A returned an empty table due to a query on \texttt{status LIKE '\%pending\%'}, which does not exist in the schema. &
eSapiens also failed initially but recovered via an introspection loop that retrieved valid status codes and regenerated the query. \\[4pt]

Unit conversion (metre $\rightarrow$ mile) &
All baselines returned the raw \texttt{distance} column, ignoring the unit conversion. &
eSapiens injected \texttt{distance/1609.34} into SQL to compute and return results in miles. \\[4pt]

Fuzzy genre match (“hip-hop”) &
Agents B and C required exact matches (e.g., \texttt{'Hip Hop'}), resulting in zero rows. &
eSapiens used a retry strategy with \texttt{ILIKE '\%hip\%hop\%'} to successfully match variants. \\[4pt]

Future-dated rows in “last-3-months” query &
Baselines included future timestamps without warnings. &
eSapiens flagged the anomaly and regenerated SQL with a bounded date predicate. \\[4pt]

Multi-step request (top albums + buyer country) &
Agent-B misinterpreted the join; Agent-C declined to respond. &
eSapiens parsed and executed a two-step SQL query, merging both result sets correctly. \\
\bottomrule
\end{tabularx}
\end{table*}

\end{document}